1 Article

# The statistical mechanics of community assembly and species distribution

6 *Colleen K. Kelly[a,1], Stephen J. Blundell[2], Michael G. Bowler[2], Gordon A. Fox[3], Paul H. Harvey[1],*
7 *Mark R. Lomas[4] and F. Ian Woodward[4].*

9 [1] Dept of Zoology, University of Oxford, South Parks Road, Oxford OX1 3PS, UK; Email:
10 *colleen.kelly@zoo.ox.ac.uk*

11 [2] Dept of Physics, University of Oxford, Keble Road, Oxford OX1 3RH, UK; Email:
12 *s.blundell@physics.ox.ac.uk*; *michael.bowler@physics.ox.ac.uk*

13 [3] Dept of Biology, University of South Florida, 4202 E. Fowler Ave, Tampa, FL, USA 33620;
14 Email: *gfox@mail.cas.usf.edu*

15 [4] Dept of Animal and Plant Sciences, University of Sheffield, Sheffield S10 2TN. Email:
16 *m.r.lomas@sheffield.ac.uk*; *f.i.woodward@sheffield.ac.uk*

18 **Key words**: biotic resistance; diversity and productivity; niche dynamics; idiosyncrasy;
19 neutrality; naturalized species

20 **Supplementary material**      Supplementary Table 1.
21                               Appendix A: Dispersal
22                                        Failure of models involving attenuation with distance.
23                                        A single step approach.
24                                        Figure A1.  Patterns expected from a diffusion process.
25                               Appendix B: Maximization subject to constraints and determination
26                                        of parameters
27                               Appendix C: The nature of equilibrium
28                               Appendix D: Statistical mechanics in ecology.
29                                        Statistical mechanics, maximum entropy and ecological
30                                         guilds
31                                        Maximum entropy, priors and alien species
32

33 [a] To whom correspondence should be addressed

34

34    **Abstract.** Theoretically, communities at or near their equilibrium species number resist

35    entry of new species. Such 'biotic resistance' recently has been questioned because of successful

36    entry of alien species into diverse natural communities.  Data on 10,409 naturalizations of 5350

37    plant species over 16 sites dispersed globally show exponential distributions for both species over

38    sites and sites over number of species shared.  These exponentials signal a statistical mechanics

39    of species distribution, assuming two conditions. First, species and sites are equivalent, either

40    identical ('neutral'), or so complex that the chance a species is in the right place at the right time

41    is vanishingly small ('idiosyncratic'); the range of species and sites in our data disallows a neutral

42    explanation. Secondly, the total number of naturalisations is fixed in any era by a 'regulator.'

43    Previous correlation of species naturalization rates with net primary productivity over time

44    suggests that regulator is related to productivity.  We conclude that biotic resistance is a moving

45    ceiling, with resistance controlled by productivity.  The general observation that the majority of

46    species occur naturally at only a few sites but only a few at many now has a quantitative

47    [exponential] character, offering the study of species' distributions a previously unavailable rigor.

48

48      **Introduction**

49      The effects, accidental or otherwise, that humans may have on natural systems are a classic

50      source of insight into the fundamental processes governing those systems (Darwin 1859; Elton

51      1958).  Here we use distributions of plant species' naturalization[*] to characterize the factors

52      determining entry of new species into a standing species complement, the fundamental building

53      block of natural communities.  Ecological theory of community assembly predicts that mature

54      communities – those at or near their equilibrium species number – will resist the entry of new

55      species.  Such 'biotic resistance' is proposed to occur either through *in situ* coevolution filling all

56      available niche space, or by ecological sorting to find the combination of species best able to

57      exploit available resources.  The resulting complex matrix of interactions is supposed to leave

58      little niche space in the existing community into which a newcomer may easily insert itself, thus

59      regulating community diversity (Elton 1958; Hutchinson 1959; MacArthur 1965; May 1973;

60      Pimm 1991; Tilman 2004).

61      Biotic resistance has been interpreted on a practical level to mean that highly diverse

62      communities are protected from invasion by species not currently a part of the community, and

63      small-scale manipulations under natural conditions largely support this expectation (Levine 2000;

64      Kennedy et al. 2002).  However, two patterns of species' naturalisation at greater geographic

65      scales and incorporating longer time-spans seem to contradict these observations: regional

66      inventories of species occurrences show highly diverse communities readily invaded by

67      naturalized species (Lonsdale 1999; Stohlgren et al. 1999; Sax 2002); successful naturalizations

68      are not being offset by the concomitant extinctions of native species that would be expected if

68

[*] Here we use 'naturalized' to denote merely having an established population; 'invasive' (pest)

species are a subset of naturalized species, but not all naturalized species become pests.

69  niche filling regulates community assembly (Sax and Gaines 2003).  These observations

70  challenge the idea that complex interactions regulate the successful entry of new species into

71  natural communities, and pose the question as to what, then, determines entry of a species into a

72  community.

73      In this study we exploit the global-scale 'natural experiment' created by the escalation of

74  species naturalizations over the last century.  We employ these data to examine the large-scale

75  patterns of species naturalizations and community assembly through the high-power lens of

76  statistical mechanics.  Statistical mechanics uses probability theory to provide a framework

77  relating the properties of large numbers of individual units to the bulk properties of the whole,

78  revealing emergent properties that give insight into the regulators of individual behavior not

79  available from considering one or a few individuals independently.  Statistical mechanics

80  underlies much of the realm of the physical sciences, but also has been useful in problems such as

81  the distribution of wealth (Drăgulescu and Yakovenko 2001) and the ubiquitous lognormal

82  distribution of individuals among species in ecological communities (Pueyo et al. 2007; Dewar

83  and Porté 2008; Harte et al. 2008; Bowler and Kelly 2010).

84      Our approach has produced new insights into several fundamental ecological processes.

85  First, we have derived an analytical explanation of community assembly able to incorporate

86  naturally all the above observations.  From this, we are able to conclude that biotic resistance

87  exists, but as a moving ceiling regulated by some external factor; combining these findings with

88  earlier work (Woodward and Kelly 2008), we infer that external factor to be net primary

89  productivity (NPP) or some process innately linked to NPP.  Secondly, we have identified a

90  quantitative (exponential) character to the general observation that the majority of species are of

91  restricted distribution and only a few are widespread.  This pattern is an emergent property

92  deriving from the fundamental nature of niches themselves and does not require the operation of

93  any particular trait of any particular niche.  Lastly, the simple exponential distributions make

94  possible analytical tools carrying with them a degree of rigor not previously available to the

95  comparative study of species' distributions (see (Gotelli et al. 2009).

96  **Materials and Methods**

97  We collated data on 10,409 naturalizations of 5350 unique plant species over 16 sites

98  dispersed globally, determining the number of sites at which each unique species occurred.  We

99  also recorded the number of species in common between sites, grouping sites first into all

100  possible pairwise combinations, next into all possible triplet combinations, and finally into all

101  possible quadruplet combinations.

102  Because species naturalization is largely tabulated at the country scale, our study is at this

103  scale.  Site selection was dictated by the availability of naturalized species lists including all

104  known established alien pteridophyte, gymnosperm and angiosperm species, and not restricted to

105  invasive pests.  The 16 sites meeting these criteria and included in the study are: Chile (Castro et

106  al. 2005), Czech Republic (Pysek et al. 2002), Estonia (Anonymous 2007e), Galapagos (Tye

107  2001), Hawai'i (Wester 1992), Israel (Dafni and Heller 1990), Japan (Anonymous 2007d), Latvia

108  (Anonymous 2007c), New Zealand (Healy and Edgar 1980; Webb et al. 1988; Edgar and Connor

109  1999), Poland (Anonymous 2007b), Singapore (Corlett 1988), Swaziland (Braun), Switzerland

110  (Wittenberg 2005), Taiwan (Wu et al. 2004), United Kingdom (Preston et al. 2006), and

111  Wyoming (Anonymous 2007a).  Subspecies were subsumed under the name of their parent

112  species in determining the number of unique species.

113  In order to investigate the possible effect of dispersal on the observed distributions, we

114  performed a Mantel test of correlation between geographical distance and number of species

115  shared between pairs of sites (table A1) using the R-package MANTEL module (with 9999

116  permutations) (Casgrain and Legendre 2001).

117

117    **Observations**

118          Three important properties were revealed by our treatment of the data. First, species

119    naturalizations show an exponential distribution of the number of naturalized species $S(n)$ found

120    at $n$ sites (fig. 1). To correspond to the analyses illustrated in figs. 2 and 3, the exponential is fit to

121    $n \geq 2$ using maximum likelihood; the relationship is

122    $S(n+1) = 0.59S(n)$ or $S(n) = S_0 e^{-0.52n}$

123    where $S_0$ is 2343 and the coefficient in the exponent (- 0.52) is uniquely related to the number of

124    naturalized species summed over sites, which we term *the alien footprint*, $M_1 = \sum nS(n)$ (see

125    Appendix B).

126

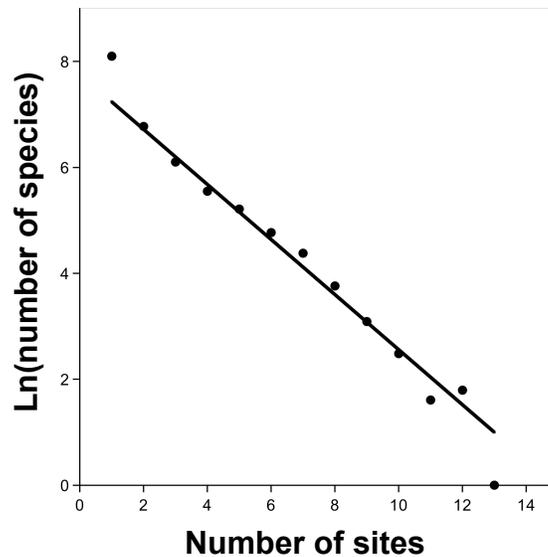

127    **Figure 1.  Number of species as a function of number of sites**. The number of naturalized species
128    $S(n)$ falls exponentially with the number of sites $n$ at which each is found. To correspond to the analyses
129    illustrated in figs. 2 and 3, the exponential is fit to $n \geq 2$ using maximum likelihood, with goodness of fit
130    assessed using the appropriate one-sample $\delta$-corrected Kolmogorov-Smirnov analysis [p > 0.20; (Khamis
131    2000)].

132          Secondly, there is no correlation between the number of naturalized species common to a

133    pair of sites and the separation of those sites (fig. 2).  The comparison of matrices of distance

134    between two sites and the number of species shared pairwise showed no relationship between the

135    two factors (p > 0.22).  Some correlation might be present for distances ≤ 5000 km, but if so, it is

136    not sufficient to affect the overall conclusion that at the global scale, the proportion of sites

137    sharing a large number of species does not depend on distance.  The number of shared species for

138    each site-pairing are given in table A1.

139

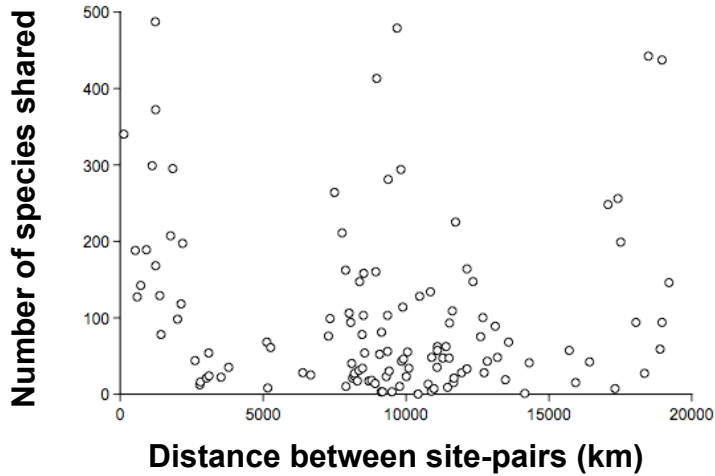

140

141    **Figure 2.  Number of species shared pairwise between sites relative to distance between sites.**
142    Distance between two sites compared to number of species shared pairwise shows no relationship
143    between the two factors.  Some correlation might be present for distances ≤ 5000 km, but the main
144    conclusion that at the global scale, the proportion of sites sharing a large number of species does not
145    depend on distance is unaffected by this possibility.  The number of shared species for each site-pairing
146    are given in table A1.

147        Finally, the number of pairs of sites sharing a given number of naturalized species falls

148    exponentially with the number of shared species (fig. 3a).  The observed distribution is an

149    exponential $y = 60e^{-0.01x}$, fitted to the individual values in fig. 2 using maximum likelihood, with

150    goodness of fit assessed using the appropriate one-sample $\delta$-corrected Kolmogorov-Smirnov

151    analysis [p > 0.20; (Khamis 2000)].  As we show below, it is also predicted from $S(n)$, assuming

152    only that the distribution is exponential.  In particular, the coefficient in the exponent is given by

153    the number of pairs (120) divided by what we refer to as the *overlap measure* $M_2 =$

154    $\sum n(n-1)S(n)/2$.  The exponentials illustrated in figs. 3b and 3c are similarly predicted.

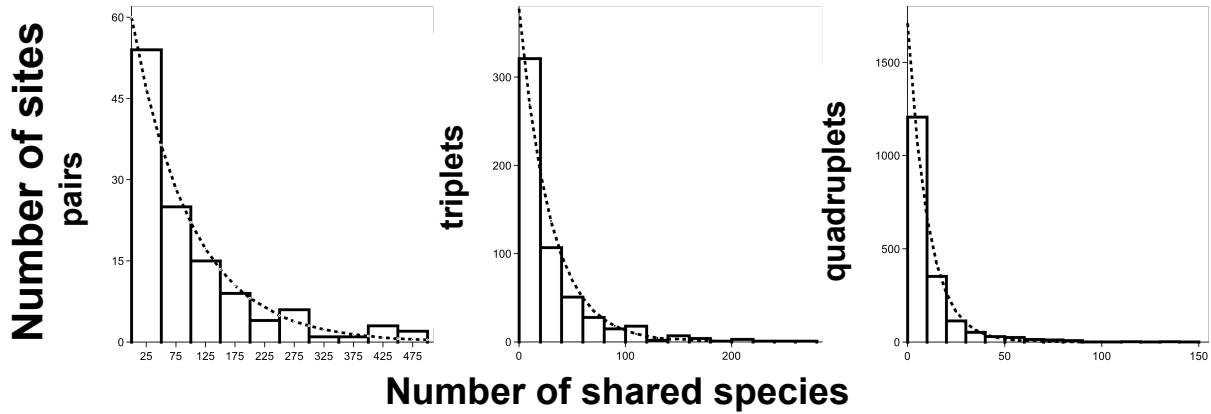

**Figure 3. Distribution of groups of sites over shared species.** In fig.3a the *y*-axis represents the number of pairs of sites sharing the number of species counted along the *x* axis. The data are binned and the exponential is calculated from *S(n)*. Figs.3b and 3c are similarly for triplets and quadruplets of sites. The ubiquity of exponentials at every level of site grouping corroborates the robustness of our findings over alternative explanations of species shared between sites (details in Appendix A: Dispersal; fig. A1).

**The Model**

No reasonable model of exponential attenuation of propagules spreading stepwise (through dispersal and transport) can produce the exponential distribution of fig. 1; even extremely contrived models are at odds with the lack of distance correlations shown in fig. 2 (see Appendix A: Dispersal; fig. A1). There is, however, an explanation that includes naturally the complexity of the biological world and this lack of correlation: the statistical behaviour of complex systems involving large numbers of components yields exponential distributions of the kind observed in figs.1 and 3. Such systems function subject to certain constraints, in this case of biological or environmental origin. The techniques of statistical mechanics are mostly employed in the physical sciences but have also found some applications in ecology (Shipley et al. 2006; Pueyo et al. 2007; Dewar and Porté 2008; Harte et al. 2008). Ecologists have long been familiar with the attempt by (MacArthur 1957; MacArthur 1960) to account for species abundance distributions in terms of a statistical model known as the 'broken stick.' MacArthur postulated that a finite resource (the stick) is partitioned at random into a given number of pieces, taken to represent species with the abundances given by the lengths. This postulate leads to an exponential distribution of species abundance as the most probable configuration subject to those constraints and can be obtained using just the techniques we discuss below. The model is not correct for

178  most species abundance distributions but does serve as model for the distribution of alien species

179  over the sites at which they are naturalized, a very different ecological problem we enlarge on

180  below.

181      The logical structure of our investigation is that we started with the hypothesis that a simple

182  argument in statistical mechanics accounts for the observed exponential distribution of species

183  over sites. We identified the necessary general conditions and constraints and found them to

184  account also for the exponential distribution of pairs of sites over numbers of species held in

185  common (fig. 3a). We were then able to predict successfully the exponential distributions shown

186  in figs. 3b and c, further supporting our hypothesis of the nature of our original observation.

187  Below, we start with the mathematical framework of our model.

188      Suppose we have $S$ objects [of so far unspecified nature] assigned to classes such that the

189  class labelled $n$ contains $s_n$ objects.  The number of ways of arranging S objects over the

190  different classes so as to achieve a configuration $\{s_n\}$, characterised by numbers in each class

191  $s_1, s_2, \ldots s_n \cdots$ is simply

192  $$W = \frac{S!}{\prod s_n!} \qquad (1)$$

193  where $\prod$ represents the continued product.

194  The quantity $W$ is proportional to the probability of finding this configuration $\{s_n\}$, provided that

195  each arrangement has equal weight; without further conditions, every object has the same

196  probability of being found in every class.  If this is not true, an additional weight factor can be

197  introduced (a *prior*) and the form of that prior is determined by the nature of the problem to be

198  addressed (Bowler and Kelly 2010; Haegeman and Etienne 2010). In statistical mechanics the

199 origin of such a factor is to be found in the dynamics of the system; this is further discussed (with

200 examples) in Appendix D.

201 ***MacArthur's broken stick***

202       To match the mathematics of Eq. (1) to reality it is necessary to specify the nature of the $S$

203 objects and the classes labelled by $n$.  In MacArthur's broken stick, the objects are species and the

204 class labelled by $n$ is the class of all species with population $n$ individuals (see also Pueyo et al

205 2007).  The environmental constraints to be applied are first that there are a given number of

206 species $S$ (the number of pieces into which the stick is to be broken) and secondly that $\sum ns_n$ is

207 fixed – this is the length of the stick; the sum of all the pieces is equal to the original length.  The

208 most probable of the configurations $\{s_n\}$ is found by maximising (1) subject to the constraints –

209 an operation which is mathematically well defined – where the constraints are on the number of

210 species and available resources to be subdivided.  The solution is

211     $s_n = s_0 \exp(-\beta n)$                                    (2)

212 (the parameters of this exponential are determined by the values of the constraints; see Appendix

213 B).  If the stick is broken randomly then the distribution of species with population $n$ as a

214 function of $n$ is exponential, provided Eq. (1) contains the essential underlying biology and the

215 constraints are the only ones that matter in this problem.  For most guilds, (2) is not an acceptable

216 species abundance distribution (Rosenzweig 1995).  The above conditions are not sufficient for

217 this problem and indeed the papers by Pueyo et al (2007), Dewar and Porte (2008) and Harte et al

218 (2008) are attempts to identify additional assumptions or constraints required to produce a log

219 series distribution, and the biological nature of such additions (see Appendix D). That particular

220 problem has been solved by Bowler and Kelly (2010).

221 ***The distribution of species over sites***

222        In our problem of the distribution of naturalized species over a number of sites, the objects

223    in Eq. (1) are naturalized species and the classes are defined by the number of sites at which a

224    species is to be found.  Thus here we identify $s_n$, general in Eq. (1), with the number of species

225    found at $n$ sites, $S(n)$.  The most probable configuration $\{s_n\}$ is obtained by maximising Eq. (1)

226    with respect to all $s_n$, subject to conditions dictated by the nature of this problem.  The first is

227    that a given number of species S is involved and the second is that the sum $\sum nS(n) = M_1$ is fixed

228    – this is the analog of the length of MacArthur's stick (and so must correspond to some fixed

229    resource, over 16 sites, to be partitioned) and it is the total number of alien establishments over

230    the 16 sites available to us.  It is of such importance that we have given it a name; the *alien*

231    *footprint* introduced earlier.  When (1) is maximised with respect to all $S(n)$ subject to these

232    constraints, the most probable distribution of species over the number of sites $n$ at which they are

233    found is given by

234    $S(n) = S_0 \exp(-\beta n)$                                                    (3)

235        The mathematical constraint on the number of naturalized establishments found in the 16

236    sites considered  $M_1 = \sum nS(n)$ , which is also the sum of site diversities, implies a biological

237    constraint.  The rate at which the exponential decreases is controlled by the *mean alien footprint*

238    for these 16 sites, $\bar{n} = \sum s_n n/S = M_1/S$, the number of sites reached averaged over all species

239    and in (3) the value of $\beta$ is determined by the value of $1/\bar{n}$ $[\beta = -\ln(1 - 1/(\bar{n}-1))]$; see fig. 1.

240    [The constants $S_0$ and $\beta$ are obtained from (3) by evaluating the sums over *S(n)* and over *nS(n)*.

241    This is discussed in greater detail in Appendix B.]

242        Thus the observed exponential in fig. 1 is reproduced by two ecological assumptions. First,

243    that the *alien footprint* has a fixed value (it is a conserved quantity) and the value is determined

244    by ecological constraints.  Secondly, the nature of the world is such that Eq. (1) is indeed

245    proportional to the probability of finding some specified configuration; there is a sense in which

246    species are equivalent.  Any other ecological forces then do not affect this distribution. The origin

247    of the ecological forces that do fix the alien footprint has not been completely established, but

248    MacArthur's idea of partitioning a limited resource is extremely suggestive. Resource availability

249    is limited – ultimately by the degree of insolation, water availability and $CO_2$ – and is reflected in

250    productivity. Species diversity has been shown to be highly correlated with net primary

251    productivity, and naturalization rates with increases in productivity over time (Woodward and

252    Kelly 2008). This is considered further in the discussion at the end of the paper.

253        There is of course nothing special about 16 sites – they were merely those for which

254    appropriate data were found. It should be clear that the conserved alien footprint defined above is

255    for that sample of 16 sites and that as time goes on the alien footprint for those particular sites

256    stays pretty much constant. If more sites were available the alien footprint for the larger sample

257    would be bigger, but would not change much with time. Thus the slope of the exponential in $n$ is

258    a function of the number of sites in the sample. It would be expected that if the fractional variable

259    $n/N$ is defined for the case of $N$ sites then the mean alien footprint *per site* would be independent

260    of $N$ and the slope of the exponential expressed as a function of $n/N$ would not depend on $N$. This

261    is not easy to test with any precision, but on selecting a random sub-sample of 8 sites from the 16

262    it is indeed the case that the mean alien footprint is halved and the mean alien footprint per site

263    remains the same. Computer simulations for the random distribution of species over sites in such

264    a way that (3) is satisfied yields the equivalent result for sub-samples of various numbers of sites.

265    Thus the mean alien footprint per site is fixed regardless of the number of sites.

266    ***The distribution of sites over species***

267        Figure 3a shows the distribution of the number of pairs of sites over the number of

268    naturalized species common to both and is again an exponential. This distribution follows from

269    the subsidiary ecological assumptions that Eq. (1) is proportional to the probability of finding

270    some specified configuration of pairs [ $s_n$ in (1)] over $n$ species in common; that there is a sense in

271    which sites are equivalent.  A constraint equivalent to the length of MacArthur's stick is wholly

272    determined by the exponential distribution $S(n)$ for $S(n)$ species over $n$ sites; it is that the sum of

273    the number of pairs with $m$ alien species in common multiplied by that number $m$ of common

274    species is constrained.  This sum is easily evaluated.  Pairs can be chosen from a set of $n$ sites in

275    $n(n-1)/2$ different ways and so a species found at $n$ sites will also be found at $n(n-1)/2$ pairs of

276    sites.  Summing over all $n$ yields a total overlap measure $M_2$ given by $M_2 = \sum n(n-1)S(n)/2$

277    which counts up all pairs of sites and sums the number of common species over all pairs and is

278    thus the required sum.  Because $S(n)$ is already determined, $M_2$ is fixed, essentially by the same

279    biological constraints that limit $M_1$.  The coefficient in the exponential in fig. 3a is given by the

280    number of pairs (120) divided by the overlap measure $M_2$.

281    Our hypotheses now allow us to predict that the distribution of triplets of sites over the

282    number of species common to all three will also be exponential and with parameters given by the

283    numbers $S(n)$ for the distribution of species over the number of sites.  The quantity

284    $M_3 = \sum n(n-1)(n-2)S(n)/6$ is the analogue of $M_2$ and is again fixed; we obtain an exponential

285    distribution with coefficient equal to the number of triplets (560) divided by $M_3$.  This is shown

286    in fig. 3b, together with the exponential for quadruplets, with a coefficient given by the number

287    of quadruplets (1820) divided by the analogous moment $M_4$, in fig. 3c.  The calculated

288    exponentials are again in agreement with the data and support our hypotheses.

289    Finally, consider the distribution of single sites over the number of naturalized species.  The

290    mean number of species per site is given by $M_1/N$ $\left(\sum nS(n)/N\right)$, the alien footprint divided by

291    the number of sites, and this singlet distribution is also exponential under the statistical

292    assumptions.  With only 16 sites the distribution is not very well defined by the data, but

293    maximum likelihood and a Kolmogorov-Smirnov test (Khamis 2000) show them to be consistent

294    with being drawn from the exponential.

295    The distributions of sites over species and of various multiplets of sites holding species in

296    common all contain information. The exponentials observed show that in every case the

297    distribution corresponds to the maximum amount of missing information (the distributions most

298    likely to be encountered) after fixing the alien footprint.

299    The ecological implications of the success of our assumptions in reproducing the observed

300    distributions are first that the overlap measures $M_n$ are fixed (already ensured by the form of the

301    distribution of species over sites) and secondly that sites are (without constraints) indifferent to

302    the classes of the number of species, pairs of sites similarly indifferent to the classes of the

303    number of species held in common and so on; Eq. (1) is applicable to all these classes.  Thus

304    these distributions imply that sites are in some sense equivalent, just as the distribution of species

305    over sites implies an equivalence of species.

306    **Discussion**

307    While it is widely observed that, in nature, species are generally restricted in distribution

308    and relatively few species are widely distributed (Pielou 1979; Brown 1995; Gaston 2003), no

309    definitive quantitative pattern of species distribution has previously been revealed (Gotelli et al.

310    2009). The number of species $S(n)$ at $n$ sites might fall with $n$ in many different ways, and the

311    exponential observed here is new information revealing underlying processes.

312    The distribution of species naturalizations contains an analog of the mean energy term $kT$ in

313    the theory of gases in the *mean alien footprint per site*, the number of alien establishments

314    averaged over all species and all sites.  With fixed numbers of naturalized species and of sites,

315    increasing the mean number of sites per species ($\bar{n}$) dictates an increase in the average number of

316    species per site; for a given number of species distributed among a given number of sites, the sum

317    of sites over naturalized species is equal to the sum of species over sites.  For naturalized species

318    we suggest that the determinant of this fixed number of alien establishments per site, an

319　ecological analogue of thermodynamic temperature $T$, may be associated with productivity, in the

320　light of the relationship between plant species naturalization rate and increasing net primary

321　productivity (NPP) over time (Woodward and Kelly 2008).  An increase in productivity would

322　then increase this 'ecological temperature,' to produce a new most probable exponential in which

323　species are found at more sites, and more species are found per site.  This would be so regardless

324　of whether total number of naturalized species increases or not [where species do not increase,

325　the analog is heating a box of gas from outside; where species increase, an injection of hotter gas

326　into the box].

327　　　The fixed nature of the number of naturalizations per site does not imply that no further

328　naturalizations are possible; such an extreme interpretation is not necessary.  The model is not

329　likely to be perfect and the world is not likely to be in equilibrium.  There are also stochastic

330　effects with a small sample, such as only 16 sites.  Finally, we envisage the lid on the total

331　number of naturalizations being raised as global climate changes.

332　　　The relevant point is that the dynamic 'relaxes' rapidly into the (quasi) equilibrium

333　configuration, achieving a new maximum number of species within the time scale over which

334　changes in productivity occur.  Evidence for this may be found in the observed exponential itself

335　and, independently, in the close tracking of net primary productivity (NPP) by local (site)

336　naturalization rates shown in (Woodward and Kelly 2008) using a large proportion of the data

337　included here (Online Appendix C: The nature of equilibrium).  In this picture biotic resistance is

338　best portrayed as a moving ceiling responding to generally increasing productivity levels; the

339　apparent 'failure' of biotic resistance is rather a reflection of its innate character.  In such a picture

340　the current escalation of species naturalizations, carrying with it potentially destructive invasive

341　weeds (Rejmánek and Randall 2004; Ricciardi and Kipp 2008), will continue as long as NPP

342　continues to increase, a phenomenon generally attributed to ongoing global climate change and

343   potentially tied into increasing levels of atmospheric carbon dioxide (Woodward and Kelly
344   2008).

345       Our central premise in obtaining the most probable distributions by maximising Eq. (1) is
346   that, without the specified conditions, every object (species or sites or groups of sites) has the
347   same probability of being in any class.. From the observed exponential distribution of
348   naturalized species over sites, we infer that every species in our data set has the same a priori
349   probability of being in any class and all arrangements corresponding to a given configuration are
350   equally probable, similarly for sites over species. One ecological model of this would be that
351   every species is identical and further that every site is identical; the wide range of environments
352   and species comprising our data set and the reported variety of mechanisms for individual cases
353   of naturalisation (Mack et al. 2000; Mitchell et al. 2006) disallows this assumption.

354       A reasonable basis for the observed distributions and the consequent inference of
355   independence in the action of the component species and sites is provided by the concept of
356   idiosyncrasy (Pueyo et al. 2007). Idiosyncratic species each operate within the aegis of a unique,
357   highly complex niche which dictates that any species plucked at random has the same probability
358   of ending in the class characterised by that species being found at *n* sites. Like Hutchinson's
359   classic 'n-dimensional hypervolume' (Hutchinson 1957), idiosyncratic niches contain the full
360   range of factors permitting a species to persist at a site, environmental conditions, competitors,
361   consumers, infectious diseases and mutualists as well as resources. With this definition, the
362   distribution of naturalized species over the number of alien sites reached *(n)* is given by an
363   exponential once we maximise the number of equivalent configurations with Eq. (1). Similarly,
364   the distribution of (idiosyncratic) pairs of sites over classes defined by the number of species in
365   common is given by an exponential once the number of configurations is maximised. That
366   potential species (and potential niches) are so varied is the underlying assumption of the

367    idiosyncratic model of species abundance, so that 'the bits of information which are different in

368    different [ecological] models cancel out' (Pueyo et al. 2007).

369        Previous applications of statistical mechanics to community assembly have focused on the

370    lognormal distribution of individuals over species within a guild (ecologically similar taxa) of a

371    single community, and so have not had information necessary to discriminate between neutral

372    and idiosyncratic explanations [although recent analyses have demonstrated that even highly

373    similar co-occurring species cannot be assumed to meet the fundamental neutrality criterion of

374    species interchangeability (Kelly et al. 2008; Leibold 2008; Kelly et al. 2010). The relation

375    between our treatment of naturalized species, (Pueyo et al. 2007), and other recent works

376    employing statistical mechanics in ecology (Dewar and Porté 2008; Harte et al. 2008; Bowler and

377    Kelly 2010) is discussed in Online Appendix D: Statistical mechanics in ecology.

378        Independent evidence ties our findings directly into the fundamental nature of community

379    assembly: free-living heteroflagellate communities show a similar exponential distribution of

380    species across sites (Patterson 2003), as do tree species from the tropical deciduous forest of

381    México (Trejo and Dirzo 2002). This is not particularly surprising: the relationship between

382    productivity and diversity in naturalized species reported in Woodward and Kelly (2008)

383    suggests general correlation of species diversity with productivity and the determinants of

384    productivity (Mittelbach et al. 2001; Hawkins et al. 2003; Gillman and Wright 2006; Kreft and

385    Jetz 2007). The natural inference is of a similarly general directionality between productivity and

386    diversity, an inference in accord with recent theoretical treatments relating diversity to both

387    complexity and productivity (Tokita 2004; Tokita 2006; Dewar and Porté 2008; Harte et al.

388    2008). At smaller scales, the reverse has been observed, with productivity apparently causally

389    affected by diversity (Flombaum and Sala 2008). Scale-dependence in the directionality of the

390    relationship is an intuitively satisfying integration of these differences, with productivity

391    determining the population process of species entry as proposed in Tilman (2004), along the

392     major axis of the relationship, and filtering of species (sampling) through subsequent species

393     interactions affecting the variation at any particular point along that axis as in Flombaum and

394     Sala (2008).

395          In conclusion, the primary result of our treatment of species naturalization is a new window

396     on the fundamental processes governing community assembly and diversity – identifying the

397     significance of the alien footprint, the implications of a causative role for productivity and the

398     rapidity with which equilibrium species number can be reached – but it also generates subsidiary

399     insights.  Regardless of the extent to which an assumption of idiosyncrasy holds, the data of figs.

400     1-3 make it most unlikely that any single pronounced signature will reveal species that can easily

401     naturalize; while there may be geographically or taxonomically local generalities, no one solution

402     will be universal, consistent with recent reviews of empirical species' naturalization studies

403     (Mack et al. 2000; Mitchell et al. 2006).  The implication of species idiosyncrasy also provides an

404     explanation in the same vein for the general observation that the majority of species have a

405     restricted distribution and few species are widespread over many sites; this pattern is an emergent

406     property deriving from the fundamental nature of niches themselves, and does not require the

407     operation of any particular trait of any specific niche (*cf.* (Brown 1984; Brown 1995).  We have

408     shown here a quantitative [exponential] character to that general observation, making possible an

409     analytical tool carrying with it a degree of rigor not previously available to the comparative study

410     of species' distributions.

**Acknowledgments**

412          We thank Oliver Pybus for his comments on an earlier draft of the manuscript, Chris

413     Preston for providing the UK Alien Species list and David Patterson for giving us access to his

414     heteroflagellate data.  CKK thanks the Chamela Biological Station for its hospitality during the

415     manuscript development.

416

**Supplementary Table 1. Number of species shared between sites.** Authorities can be found in Literature Cited section of the main text.

| | Chile | Czech Republic | Estonia | Gala-pagos | Hawaii | Israel | Japan | Latvia | New Zealand | Poland | Singapore | Swaziland | Switzerland | Taiwan | UK |
|---|---|---|---|---|---|---|---|---|---|---|---|---|---|---|---|
| Chile | 0 | 225 | | | | | | | | | | | | | |
| Czech Republic | 225 | 0 | | | | | | | | | | | | | |
| Estonia | 100 | 372 | 0 | | | | | | | | | | | | |
| Galapagos | 22 | 47 | 21 | 0 | | | | | | | | | | | |
| Hawai'i | 128 | 147 | 62 | 99 | 0 | | | | | | | | | | |
| Israel | 28 | 44 | 24 | 19 | 41 | 0 | | | | | | | | | |
| Japan | 248 | 413 | 211 | 68 | 25 | 52 | 0 | | | | | | | | |
| Latvia | 75 | 299 | 340 | 15 | 47 | 21 | 162 | 0 | | | | | | | |
| New Zealand | 294 | 442 | 256 | 55 | 264 | 42 | 479 | 199 | 0 | | | | | | |
| Poland | 33 | 188 | 142 | 9 | 28 | 12 | 78 | 127 | 94 | 0 | | | | | |
| Singapore | 7 | 10 | 3 | 27 | 57 | 10 | 61 | 3 | 18 | 3 | 0 | | | | |
| Swaziland | 43 | 54 | 30 | 48 | 94 | 28 | 89 | 23 | 93 | 14 | 17 | 0 | | | |
| Switzerland | 48 | 189 | 118 | 13 | 43 | 16 | 114 | 98 | 146 | 78 | 0 | 21 | 0 | | |
| Taiwan | 59 | 81 | 40 | 57 | 158 | 31 | 197 | 24 | 103 | 17 | 54 | 62 | 23 | 0 | |
| UK | 134 | 487 | 295 | 34 | 109 | 35 | 281 | 207 | 437 | 129 | 4 | 56 | 168 | 46 | 0 |
| Wyoming | 103 | 147 | 106 | 8 | 68 | 7 | 160 | 94 | 164 | 27 | 1 | 15 | 34 | 35 | 76 |

**Appendix A. Dispersal**

*Failure of models involving attenuation with distance.*

It is natural to think of dispersal by diffusion as limiting the distance travelled by propagules and hence the number of sites reached. For a given diffusion parameter, the probability of a propagule ending up a certain distance $R$ from the source is a normal distribution. In two dimensions were all sites (or a fixed proportion thereof) within a circle of radius $R$ to be captured by a species ending up at $R$, then an exponential as in Figure 1 would in fact be generated, but the number of pairs of sites sharing a species would decrease rapidly with the separation of the members of the pair. In fact the idea that all sites within a circle of radius $R$ be captured is a red herring. In a random walk the number of sites visited on average grows almost linearly with the number of steps allowed (which corresponds to the diffusion parameter) and many of these sites are outside the circle possessing the final radius, the average value of which grows only with the square root of the number of steps. Fluctuations about the mean number of sites visited for a fixed number of steps cannot generate the desired exponential. Dispersal by diffusion is incapable of reproducing the data of fig. 1 (unless the number of steps allowed is exponentially distributed; see below), and cannot generate anything resembling fig. 2.

As an alternative, consider a process in which propagule drift is all one way and accidents attenuate the flux of propagules exponentially with distance. Each site a propagule passes is adopted but once a propagule suffers an accident it goes no further. This model is not realistic but was contrived to generate an exponential dependence (as in fig. 1) on $n$ of the probability of $n$ sites being taken. This it will do, provided the probability of getting from site $n$ to site $n+1$ is some universal constant $x$ (in an explicitly spatial picture this would correspond to successive sites having a fixed separation). Thus the probability of reaching site $n+1$ is conditional on the probability of site $n$ having already been reached and an equation of the form given in the caption to fig.1 results [where the value of $x$ is 0.59]. Because very few propagules make – say – 15 steps the probabilities of distances between species being above 10 units are very small in comparison with the probabilities of gaps of a few units only; quite unlike fig. 2. The number of species found at two sites separated by distance $d$ inevitably falls with $d$. It is easily shown that the number of pairs of sites separated by $d$ (integer) units falls as the factor $x^d$ – an exponential.

It is not necessary in the above scenario for propagules to travel in straight lines, merely that each link is of constant length. An illustration of the problem is provided by fig. A1a, b. To make these figures we used a combination of random walk in two dimensions and attenuation, but attenuation is the driving feature. Species were launched on a two dimensional grid and executed a random walk of $D$ steps, where $D$ was chosen at random for each species from the same exponential probability distribution. For each species any site visited once or more was counted as taken; the origin could be crossed but not taken (because there the species would not be alien). Fig. A1a displays a good exponential – which it should because the model was designed to do just that. The distribution of pairs of sites as a function of their separation was calculated and the analogue of fig. 2 is shown in fig. A1b, quite unlike fig. 2. In both there are pairs of sites with several hundred species in common, but these are only closely separated in fig. A1b. The clusters of points at separations given by the lengths of the hypotenuse of right angled triangles with two integer sides is a reflection of the geography of our unrealistic two dimensional world, just as the vertical stripes in fig 2 reflect the geography of our planet. The observations summarised in figs.

596     1 and 2 can hardly have been generated by any kind of dispersal mechanism involving decreasing
597     probability with distance.

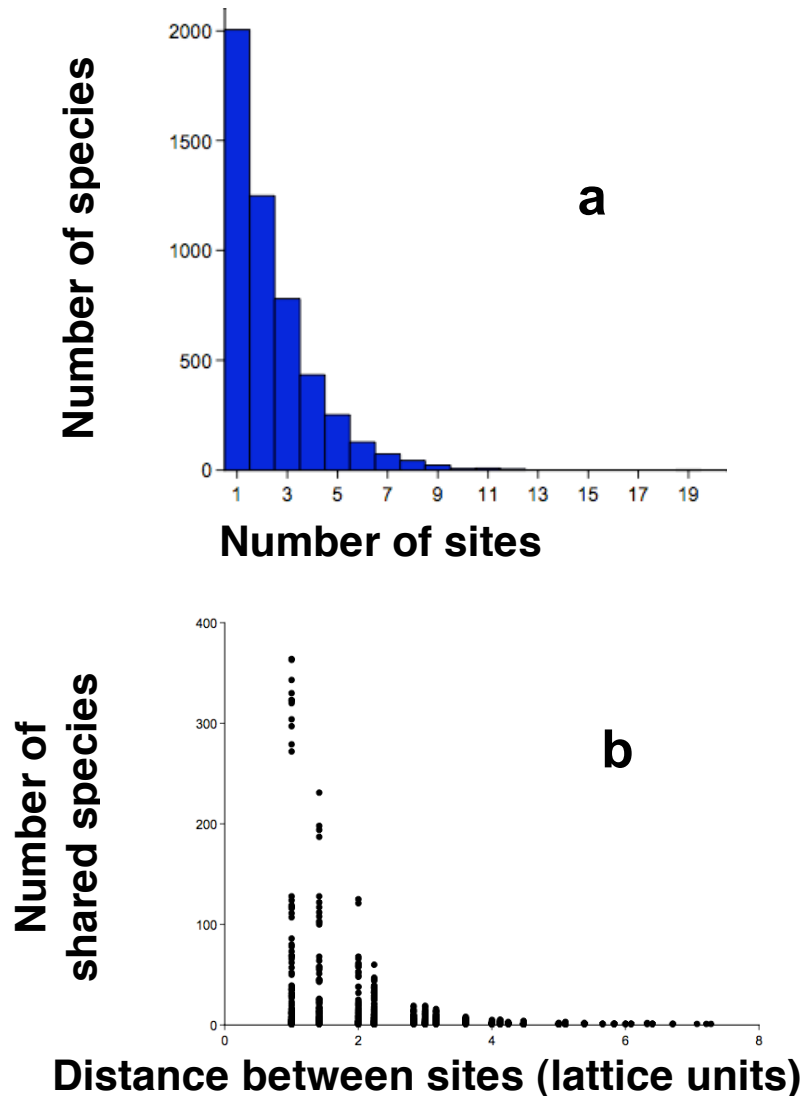

598
599
600
601

602
603
604
605 Figure A1.  Patterns expected from a diffusion process.   a. Distribution of number of species as a function
606 of number of sites.  The width of the exponential is set by a single species found at 19 sites.  b. Number of
607 species shared pairwise between sites relative to distance between sites.  Compare with the actual data
608 shown in fig. 2 in the text.

609     To generate an exponential distribution of species with the number of sites reached the spread
610 must proceed stepwise, with a single probability of each step or link in the propagation chain
611 failing. It is not necessary that each step is over the same distance and if this is not the case the
612 lack of correlation in fig. 2 may be less of an objection.  However, abandoning that assumption
613 does not increase the plausibility of this already highly contrived model. If species can easily
614 reach sites at antipodes (as they can; fig. 2, table A1) and steps can be very long, propagation by
615 a number of sequential steps is even more artificial. (Wyoming and New Zealand were both
616 colonised from Western Europe; NZ was not colonised by settlers from Wyoming.) It is more

617 realistic to consider spread of a species from its homeland to sites at which it is an alien in a
618 number of independent steps, perhaps along trade routes ancient and modern.

619 ***A single step approach.***
620 In a picture with independent single steps, species are launched a number of times and have a
621 certain fixed probability of hitting a target (an alien site at which a species becomes established).
622 The simplest version, that probability would be independent of species and of site, will not
623 produce the simple pattern observed. This is again a mechanistic model, but distances do not
624 enter explicitly. After a possibly large number of attempts, let the chance of a single species
625 having achieved one or more hits on any given target be $p$, with $N$ target sites. Then the
626 probabilities of a species occupying 1, 2, 3, …$n$ sites are given by successive terms in the
627 binomial expansion

628
$$P(n) = p^n \left(1 - p\right)^{N-n} \frac{N!}{(N-n)!\,n!} \qquad (A1)$$

629 and the features of this distribution are entirely different from the exponential in $n$ which is
630 observed. The model described in this section is ludicrously simple because the same probability
631 $p$ has been taken for each species and every target. If instead different values are possible for
632 each species on every site the possibilities are enormously increased and this of itself suggests a
633 statistical approach, necessarily involving the complexity of the biological world. Simple
634 mechanistic explanations do not lead to any acceptable explanations of our observations.

635 **Appendix B.  Maximization subject to constraints and determination of parameters**

636 The problem we have is to maximise the weight given by eq (1) of the text, subject to constraints.
637 This outline of the general case may help the reader to perceive the analogy between our
638 ecological problem and the statistical mechanics of gases. The function to be maximised is

639
$$W = \frac{S!}{\prod s_n!}$$

640 with respect to all $s_n$, subject to constraints. The first constraint simply imposes the condition that
641 we are working with a fixed number of objects, be they atoms or species. This condition is

642
$$G\left(s_n\right) = \sum s_n - S = 0$$

643 This is a zero order moment; the next condition is a first order moment. If the objects labelled by
644 $n$ have some attribute which we denote generally by $A_n$, then a second constraint which might
645 apply is

646
$$H\left(s_n\right) = \sum s_n A_n - A = 0$$

647 These two constraints would determine the average value of the attribute $A$. In the statistical
648 mechanics of gases $A$ is the total energy of the $S$ atoms. (Higher order moments can be introduced
649 as constraints in the same way; we do not need to go further.)

650   It is convenient to maximise the function

651   $$F(s_n) = \ln(W)$$

652   rather than $W$ itself and the necessary condition to be satisfied for all $s_n$ is

653   $$\frac{\partial F}{\partial s_n} + \lambda \frac{\partial G}{\partial s_n} + \mu \frac{\partial H}{\partial s_n} = 0$$

654   The quantities $\lambda$ and $\mu$ are at this stage undetermined but for non zero values impose the
655   constraints. An elegant explanation of the principles behind the use of these Lagrangian
656   multipliers may be found in appendix C.13 of Blundell and Blundell (2006).

657   Expanding the logarithms of the factorial functions using Stirling's theorem, the condition for an
658   extremum under constraints becomes

659   $$\frac{\partial F}{\partial s_n} + \lambda + \mu A_n = 0 \qquad \text{whence} \qquad \ln s_n = \lambda + \mu A_n$$

660   and the exponential dependence of $s_n$ on the attribute $A_n$ follows.

661   In the statistical mechanics of gases, the number of atoms in a level of energy $E_n$ decreases
662   exponentially with that energy (the Boltzmann distribution). In the problem of the distribution of
663   species over sites the $s_n$ are the number of species $S(n)$ in a class defined by a species being at $n$
664   sites; the attribute is $n$. Thus there are 873 at 2 sites, 184 at 5 sites and so on down to 1 species at
665   13 sites.

666   If the first moment of $S(n)$, $\sum n S(n)$ is constrained and the number of complexions maximised
667   subject to this constraint, then an exponential distribution of $S(n)$ over $n$ results

668   $S(n) = S_0 \exp(-\beta n)$                                                                  (A2)

669   where $\beta$ is an undetermined multiplier. However, both the normalising constant $S_0$ and the
670   constant $\beta$ are determined by the total number of species and the value of the first moment.

671   Suppose we carry out summations from $n = 1$ to infinity – we have no information on what $S(0)$
672   might be or even how meaningful it is. Then we define

673   $$S_1 = \sum_1 S(n) \qquad\qquad F_1 = \sum_1 n S(n)$$                                      (A3)

674   For a given number of species and given values of the number of species at each number of sites,
675   these numbers can be calculated from the data without any assumption about the shape of the
676   distribution. For the data collected for alien species the numbers are respectively 5350 and 10409.

677    The mean number of sites per species is given by $\overline{n_1} = F_1/S_1$ and is, from the above numbers,
678    1.946 sites per species.

679    Now substitute the expression (A2) into eqs (A3). The sums can be calculated very simply (these
680    are essentially sums over geometric series) and the following results are mathematically exact.

681    $$S_1 = S_0 \exp(-\beta)/(1 - \exp(-\beta)) \qquad F_1 = S_0 \exp(-\beta)/(1 - \exp(-\beta))^2$$

682    and hence $\overline{n_1} = 1/(1 - \exp(-\beta))$. The data are best represented here by an exponential for $n$ greater
683    than or equal to 2 and fig. 2 is restricted to $n \geq 2$.

684    Therefore define

685    $$S_2 = \sum_2 S(n) \qquad\qquad F_2 = \sum_2 nS(n) \qquad\qquad\qquad (A4)$$

686    The numbers from the data are 2049 and 7108 respectively.  The ratio $\overline{n_2} = F_2/S_2$ is 3.47.

687    We can of course substitute (A2) into equations (A4) and calculate $\beta$ in terms of the new average
688    $\overline{n_2}$.  The calculations are again simple sums of series and the result is that $\beta = -\ln(1 - 1/(\overline{n_2} - 1))$.
689    The normalising constants are also easily calculated in terms of the sums. The exponential best
690    fitted to the data points at n=2 and greater has $\beta = 0.52$ and the constant $S_0 = 2343$.

691    **Appendix C.  The nature of equilibrium**

692    The distribution of species over the number of sites is exponential, as is the distribution of pairs
693    of sites over the number of species in common. These exponentials are the most probable
694    configurations subject to the relevant constraints. Most probable configurations correspond to the
695    notion of equilibrium; once a system is in the vicinity of this configuration it is very unlikely to
696    depart substantially from it.

697    The existence of such an equilibrium merely dictates an exponential distribution of alien species
698    over the number of sites. It does not specify which species are found at 8 sites or anything of that
699    kind – it does not even require that the same species are found at 8 sites at all times (atoms hop in
700    and out of energy levels). Still less does this equilibrium require that the populations of alien
701    species are unchanging; only the presence of a certain number of species at a certain number of
702    sites.

703    It is particularly interesting that alien species have reached configurations close to equilibrium
704    and quite quickly at that.  Probably a global equilibrium with a single global ecotemperature is an
705    oversimplification, but the data are close.  Insofar as the role of human activity is concerned, this
706    would reduce the relaxation time rather than determine the distribution.  An analogous process is
707    bringing boxes of gas at differing temperatures into better thermal contact.

708    **Appendix D.  Statistical mechanics in ecology.**
709    ***Statistical mechanics, maximum entropy and ecological guilds***

710 There have been several recent papers applying methods, similar to ours, from statistical
711 mechanics to the structure of ecological guilds. The work of (Pueyo et al. 2007) derives the
712 whole family of species abundance distributions from very few assumptions. The first is their
713 idiosyncratic assumption: that every species is different (the antithesis of the assumptions of
714 neutral models). The second is concerned with the statistical properties of large ensembles of
715 complex ecological models; that the species abundance distribution can be obtained by applying
716 the principle of maximum entropy so as to obtain a probability distribution for species abundance
717 which contains minimal information (maximum entropy). In statistical mechanics this is a very
718 likely configuration because it can be obtained in a vast number of ways – if the system explores
719 these possibilities. [The problem is initially set up in terms of combinatorics, as in eq (1) in the
720 main body of this paper. This formulation sheds particular light on the role of *a priori*
721 probabilities (see below).] An important element is that rather than the entropy of information
722 theory, a quantity called the relative entropy is maximised, which requires the introduction of a
723 'prior' (Jaynes 1968; Jaynes 2003) – this is equivalent to discarding the assumption that every
724 species has the same *a priori* probability of having any abundance. The maximisation is subject
725 to two constraints: first that the probability is normalised to unity and secondly that a mean of the
726 number of individuals $n$ is constrained [as in MacArthur's broken stick model for species
727 abundance (MacArthur 1960; Etienne and Olff 2005)]. The solution is then

728 $$P(s_n) = P_\pi(n) e^{-\beta n} \tag{A5}$$

729 where $P_\pi(n)$ is the 'prior' relative to which the entropy is maximised; an *a priori* probability that
730 must be applied before maximising the purely combinatorial weight, or maximising entropy.
731 (Pueyo et al. 2007) imposed not the commonly employed uniform prior [corresponding to both
732 MacArthur's model (MacArthur 1960) and the statistical mechanics of gases (Bowler 1982)], but
733 rather

734 $$P_\pi(n) = A/n \tag{A6}$$

735 where $A$ is a constant. The prior is chosen before the total number of individuals is specified and
736 Pueyo et al argued that this choice of prior is correct for species abundance distributions because
737 it is in a certain sense scale invariant and in consequence contains no information on the
738 geographic scale or sample size (Pueyo et al. 2007).

739 The result of the particular choice (A6) is the famous log series expression

740 $$P(s_n) = \frac{A e^{-\beta n}}{n} \tag{A7}$$

741 This choice of prior corresponds to equal intervals of log $n$ being equally probable *a priori*. If the
742 relative entropy is maximised subject to an additional condition on the mean value of log $n$ then
743 the result is

744 $$P(s_n) = \frac{A e^{-\beta n}}{n^b} \tag{A8}$$

745 and if a constraint is also applied on the average value of $\left(\log n\right)^2$ then a skewed log normal
746 distribution results. We note here that (A7) is a particular member of the family of solutions (A8),
747 as indeed is the broken stick solution of (MacArthur 1960).

748 The paper of (Dewar and Porté 2008) is similar in a number of respects. Again the relative
749 entropy is maximised, but in this case the prior is taken as

750 $$P_\pi(n) = \frac{A}{n+1} \qquad\qquad\qquad\qquad (A9)$$

751 Their motivation for this choice of prior is again that it in some sense contains the least
752 information, but their criterion is drawn from coding theory rather than scale invariance.
753 Naturally a species abundance distribution close to the log series results.

754 Finally, the recent paper of (Harte et al. 2008) applies more constraints. In addition to the number
755 of species and the total number of individuals in the guild being fixed, a measure of total
756 metabolic rate is also taken as constrained. Their treatment employs a joint probability function
757 $R(n,\varepsilon)$; the probability of a species having $n$ individuals and of an individual having energy
758 requirement $\varepsilon$. A uniform prior is assumed and the entropy maximised to yield this function.
759 Integration over the continuous variable $\varepsilon$ then results in a log series species abundance
760 distribution.

761 ***Maximum entropy, priors and alien species***
762 The principle of maximum entropy is used in more than one way, mathematically equivalent but
763 different in interpretation. If all that is known of a function is the values of certain moments, then
764 maximising the entropy subject to these constraints minimises the information contained in the
765 resulting function, thus yielding the least biased estimate of the probability distribution consistent
766 with limited information. In the statistical mechanics of gases the problem is different. The
767 number of atoms in a box is known to be constrained by physics (impermeable walls) and the
768 mean energy is known to be constrained (adiabatic walls and conservation of energy).
769 Application of maximum entropy yields the most probable distribution consistent with these
770 constraints (the Boltzmann distribution). Here the constraints are real as opposed to being the
771 result of inadequate information.

772 The role of a prior distribution is clear enough when it is used to incorporate already existing
773 information and need not be mysterious. In the kinetic theory of gases each state with energy $E_n$
774 has occupation given by eq (2) of the text, but if there are $g_n$ different quantum states with this
775 same energy, then the number of atoms with energy $E_n$ is eq (2) multiplied by the *degeneracy*
776 *factor* $g_n$. The degeneracy factor plays the role of a prior [see for example (Bowler 1982)] and in
777 this context pre-dates information theory. (Pueyo et al. 2007) and (Dewar and Porté 2008)
778 appealed to principles of minimum knowledge for their priors (Pueyo et al. 2007), but in
779 statistical mechanics one expects to see machinery driving the choice of prior. We note that the
780 model of Harte et al (2008) contains a concrete realisation equivalent to the prior of Pueyo et al
781 (2007). The complexity model of (Tokita 2004; Tokita 2006) is rather successful at producing
782 species abundance distributions and the last equation on page 122 of (Hubbell 2001) contains
783 factors which are identical to the prior of Pueyo et al. (2007), yet reached from a very different

784 argument. The origin of the prior of Pueyo et al (2007) has in fact been traced to a specific piece
785 of biological machinery. This is the fundamental biological processes of the birth and death of
786 individuals, so that a species with $n$ individuals exits that class at a rate proportional to $n$ (Bowler
787 and Kelly 2010). Thus the prior of Pueyo et al (2007) is correct for species abundance
788 distributions but need not apply to the distribution of alien species, a very different ecological
789 problem.

790 In our application of statistical mechanics to naturalised species, we know not only the mean
791 value $\bar{n}$ but also that the distribution is exponential. We do not need to estimate a distribution by
792 using maximum entropy. All we need is the knowledge that with a uniform prior [the weight in
793 eq (1) correctly representing probability] an exponential results if that mean value is determined
794 by the properties of the world in which these species find themselves. This is in accord with the
795 absence of any identifiable mechanism that could bias the probabilities of the $s_n$ *a priori*.

796 **Literature cited in appendices**